\documentclass[cameraready]{Interspeech}

\title{Probing Spatial Structure in Pretrained Audio Representations}

\author[affiliation={1}, orcid=0009-0002-3839-8539, equalcontribution]{Chuyang}{Chen}
\author[affiliation={1}, orcid=0009-0003-0706-6117, equalcontribution]{Sivan}{Ding}
\author[affiliation={1}, orcid=0009-0003-3865-6910, equalcontribution]{Adrian S.}{Roman}
\author[affiliation={1}, orcid=0000-0001-8561-5204,]{Juan P.}{Bello}

\address{
    $^1$ Music and Audio Research Laboratory, New York University, USA
}

\email{chuyang.chen@nyu.edu, sd4839@nyu.edu, asr9618@nyu.edu, jpbello@nyu.edu}


\usepackage{comment}

\usepackage{booktabs}
\usepackage{multirow}
\usepackage{colortbl}
\usepackage{xcolor}
\usepackage{siunitx}
\usepackage{dblfloatfix}
\usepackage{placeins}   
\usepackage{makecell}
\usepackage{tabularx}
\usepackage{array}
\usepackage{graphicx}
\definecolor{lightgray}{gray}{0.92}
\definecolor{rank1}{RGB}{33,  113, 181}   
\definecolor{rank2}{RGB}{107, 174, 214}   
\definecolor{rank3}{RGB}{198, 219, 239}   

\newcommand{\cmt}[1]{}

\begin{document}

\maketitle

\begin{abstract}
Pretrained spatial audio encoders are increasingly used as general-purpose representations for perceptual tasks, yet their spatial encoding capabilities remain poorly understood. We introduce the Spatial Audio Representation Learning (\textbf{SARL}) benchmark, a controlled framework for evaluating spatial information in pretrained audio models. SARL probes source-level factors (azimuth, elevation, distance, class) and room-level factors (RT60, volume, shape). Experiments across diverse encoders reveal three patterns: input configuration and training paradigm shape spatial encoding; source factors are consistently easier to decode than room factors; and sensitivity analysis under controlled perturbations shows heterogeneous responses to source and room variation. These results reveal systematic biases in current pretrained audio representations. SARL is released as an open-source benchmark for reproducible evaluation of spatial audio representations\footnote{Code and datasets are available at \url{https://github.com/chuyangchencd/SARL}.}.
\end{abstract}

\section{Introduction}

Spatial audio plays a central role in immersive media, robotics, embodied AI, and acoustic scene understanding~\cite{chen2020soundspaces, gan2020look, morgado2020learning}. 
Unlike monaural recordings, multichannel signals encode directional, distance, and room-dependent cues that support three-dimensional perception~\cite{blauert1997spatial}. 
Recent advances in spatially-aware audio models have led to increasingly powerful representations learned from binaural and ambisonic recordings~\cite{zheng2024bat, yuksel2025gram}. 
However, evaluation of these models remains largely pipeline-dependent and end-to-end, making it difficult to isolate how spatial information is encoded in the learned representations.

Existing audio representation benchmarks such as HEAR~\cite{turian2022hear}, SUPERB~\cite{yang21c_interspeech}, X-ARES~\cite{zhang2025x}, and MARBLE~\cite{yuan2023marble} focus primarily on monaural signals and semantic tasks, providing limited coverage of spatial information. 
Conversely, spatial audio and speech benchmarks such as the DCASE multichannel challenges~\cite{politis2020overview}, LOCATA~\cite{evers2020locata}, NatHEAR~\cite{yuksel2025gram}, CHiME~\cite{barker2015third}, and REVERB~\cite{kinoshita2016summary} evaluate localization, enhancement, dereverberation, or recognition performance in multichannel settings. 
While valuable, these task-centric evaluations measure end-to-end system performance and therefore conflate representation quality with downstream architectures, supervision signals, and training strategies. 
Moreover, they offer limited control over the spatial factors being evaluated, making it difficult to disentangle how different aspects of the acoustic scene are represented.

We argue that spatial representation evaluation should move beyond task-centric benchmarking toward \emph{probing-based} analysis. 
Rather than measuring downstream accuracy, probing evaluates which attributes are decodable from frozen embeddings using lightweight classifiers trained on fixed representations~\cite{alain2016understanding, hewitt2019designing}. 
This paradigm enables architecture-agnostic comparison across pretrained encoders while isolating representation quality from downstream model design.

To enable such controlled analysis, we introduce the \textbf{Spatial Audio Representation Learning (SARL)} benchmark. 
SARL evaluates spatial audio representations across source-level and room-level factors using a unified linear probing protocol. 
Through simulation of spatial acoustic scenes, SARL constructs a balanced dataset with controlled variation of source and room factors, enabling systematic analysis of how spatial cues are encoded in learned representations.

Our contributions are three-fold:
\begin{itemize}
    \item A controlled simulation-based spatial audio dataset with balanced variation of source- and room-level factors.
    \item A unified linear probing protocol for architecture-agnostic evaluation of pretrained spatial audio representations.
    \item A systematic evaluation of diverse pretrained models, revealing consistent trends in spatial representation learning.
\end{itemize}

\section{Related Work}

Spatial audio modeling has evolved from task-specific localization systems toward broader spatial representation learning. 
Early work on sound event localization and detection (SELD) introduced neural architectures for jointly predicting sound classes and spatial positions from multichannel recordings~\cite{adavanne2018sound, cao2021improved}. 
Subsequent research explored richer spatial reasoning objectives and multimodal supervision for learning spatially-aware embeddings~\cite{morgado2020learning, zheng2024bat}. 
More recent approaches adopt self-supervised and masked modeling frameworks to learn spatial audio representations directly from multichannel signals~\cite{yuksel2025gram, yuksel2025wavjepa}. 
In parallel, neural audio compression models have been extended to multichannel and binaural formats, learning compact spatially structured latent representations for efficient encoding and generation~\cite{defossez2022high, ratnarajah2025banc}. 
These developments reflect growing interest in spatially-aware audio representations across both discriminative and generative paradigms.

Spatial audio models have been evaluated primarily through task-specific benchmarks. 
Datasets such as the DCASE multichannel challenges~\cite{politis2020overview}, LOCATA~\cite{evers2020locata}, CHiME~\cite{barker2015third}, and REVERB~\cite{kinoshita2016summary} assess localization, enhancement, dereverberation or recognition performance using microphone arrays. 
While these benchmarks have driven substantial progress, they evaluate end-to-end system accuracy and therefore entangle representation quality with downstream architectures and training strategies. 
More broadly, representation learning research has introduced probing methodologies to analyze which attributes are recoverable from frozen embeddings~\cite{alain2016understanding, hewitt2019designing, plachouras2025towards}. 
Although probing has become a common tool for studying learned representations in other domains, a unified probing framework tailored to spatial audio remains largely unexplored.

\begin{table}[!t]
\centering
\scriptsize
\caption{Pretrained audio encoders evaluated in SARL. Models are organized by input format and summarized by training paradigm and primary learning objective.}
\label{tab:spatial_models}
\setlength{\tabcolsep}{3pt}
\renewcommand{\arraystretch}{1.05}

\resizebox{\columnwidth}{!}{%
\begin{tabular}{llll}
\toprule
\textbf{Input} & \textbf{Model} & \textbf{Training Paradigm} & \textbf{Primary Objective} \\
\midrule

\multirow{1}{*}{\textbf{Mono}}
& A-MAE~\cite{huang2022masked}
& Self-supervised
& Masked spec. recon. \\

\midrule

\multirow{3}{*}{\textbf{Stereo}}
& SELD-S~\cite{shimada2025stereo}
& Supervised
& Joint SED + DOA prediction \\
& EnCodec~\cite{defossez2022high}
& Codec
& Neural audio compression \\
& SR-VAE~\cite{saito2025soundreactor}
& Codec
& VAE recon. \\

\midrule

\multirow{5}{*}{\textbf{Binaural}}
& BANC~\cite{ratnarajah2025banc}
& Codec
& Neural speech compression \\
& GRAM-B~\cite{yuksel2025gram}
& Self-supervised
& Masked spec. recon. \\
& S-AST~\cite{zheng2024bat}
& Supervised
& Multi-task event + localization \\
& SFD~\cite{bovbjerg2025learning}
& Self-supervised
& Spatial feature distillation \\
& W-JEPA~\cite{yuksel2025wavjepa}
& Self-supervised
& Joint-embedding prediction \\

\midrule

\multirow{4}{*}{\textbf{FOA}}
& AVSA~\cite{morgado2020learning}
& Self-supervised
& A--V contrastive alignment \\
& EINv2~\cite{cao2021improved}
& Supervised
& Track-wise SED + DOA \\
& SELD-F~\cite{adavanne2018sound}
& Supervised
& Joint SED + DOA prediction \\
& GRAM-F~\cite{yuksel2025gram}
& Self-supervised
& Masked spec. recon. \\

\bottomrule
\end{tabular}%
}
\end{table}

\section{Methodology}

We evaluate pretrained audio encoders using a controlled probing framework to determine whether spatial factors are encoded in frozen representations. The benchmark contains seven tasks covering source-level factors (azimuth, elevation, distance, event) and room-level factors (RT60, volume, shape). Spatial scenes are synthesized with independent control over each factor. Models are evaluated with frozen backbones and linear probes. In addition, we measure representation sensitivity under controlled source and room perturbations.

\subsection{Model Selection}

We evaluate pretrained audio encoders spanning two design axes: input format (mono, stereo, binaural, or first-order Ambisonics) and training paradigm (self-supervised, supervised, or codec-based) (Table~\ref{tab:spatial_models}). These differences expose models to distinct spatial cues and learning signals, enabling comparison across diverse representation learning strategies.

The evaluated models include self-supervised encoders AudioMAE (A-MAE), GRAM (GRAM-B for binaural and GRAM-F for FOA), wav-JEPA (W-JEPA), SFD, and AVSA; supervised sound event localization and detection models SELDnet (SELD-S for stereo and SELD-F for FOA), EINv2, and Spatial-AST (S-AST); and codec-based encoders SoundReactor-VAE (SR-VAE), EnCodec, and BANC. We use these shorthand names throughout the paper.

\subsection{Data Generation}

Single-source spatial scenes are synthesized to provide controlled coverage of spatial factors. Source clips are drawn from a balanced 7-class pool constructed from ESC-50~\cite{piczak2015esc}, MUSAN~\cite{snyder2015musan}, and UrbanSound8K~\cite{salamon2014dataset}, with 4,000 clips per event class split into 3,200/400/400 train/validation/test examples. Clips are converted to mono, resampled to $24 \mathrm{kHz}$, RMS-normalized to $-24 \mathrm{dBFS}$, and padded or cropped to 10s.

Source-level tasks use RIRs rendered with AudibleLight~\cite{cheston2025audiblelight} on Gibson meshes~\cite{xia2018gibson}, sampling azimuth in $[-180^\circ,180^\circ]$, elevation in $[-60^\circ,60^\circ]$, and distance in $[0.5,2.5],\mathrm{m}$. We employ AudibleLight for these tasks because its ray-traced propagation on realistic room meshes provides geometrically faithful spatial cues, making it well suited for evaluating localization-related factors. Room-level tasks use RIRs generated with PyRoomAcoustics~\cite{scheibler2018pyroomacoustics}, sampling RT60 in $[0.1,3.0],\mathrm{s}$, room volume log-uniformly in $[60,2500],\mathrm{m}^3$, and room shape across four classes (cube, flat, corridor, rectangular). We use PyRoomAcoustics for room-level tasks because it enables independent control and uniform sampling of RT60, volume, and room shape, which is not possible with realistic room meshes. Each pipeline produces $12{,}000/3{,}600/3{,}600$ train/validation/test RIRs with rooms disjoint across splits.

Final scenes are rendered by pairing dry signals and RIRs within each split and convolving them on-the-fly, ensuring both sources and rooms remain disjoint across train, validation, and test sets. Sampling is deterministically seeded for reproducibility and spatial factors are drawn to approximate uniform label distributions. For evaluation consistency, we render 15,000/3,000/3,000 train/validation/test scenes per epoch within each task family. Each scene is rendered identically in stereo, binaural, and FOA to enable cross-format comparison under matched acoustic conditions.

\subsection{Evaluation Protocol}

We evaluate seven probing tasks spanning source-level and room-level factors. Continuous factors are discretized into linearly spaced bins: azimuth (36), elevation (12), distance (20), and RT60 (29). Categorical factors include class (7 classes), volume (5 logarithmic bins), and shape (4 classes).

Audio input is resampled to each encoder’s native sampling rate.
Frame-level embeddings or token sequences are mean-pooled to obtain scene-level representations. For each task, a linear classifier is trained with cross-entropy using Adam ($lr=1\times10^{-4}$) and cosine decay for 20 epochs. Discretized continuous factors use Gaussian soft labels centered at the ground-truth bin, while categorical factors use one-hot targets.

Continuous factors are evaluated using normalized mean absolute error (MAE). For ground-truth value $y$ and predicted bin center $\hat{y}$, the error is $|y-\hat{y}|$. Averaging over the evaluation set yields MAE, and performance scores are computed as $1-\mathrm{MAE}/R$, where $R$ denotes the full range of the factor. Categorical factors are evaluated using macro-F1. We report results averaged over three random seeds.

To aggregate results across tasks, we apply baseline normalization
$\phi(x; b) = (x-b)/(1-b)$,
where $b$ denotes the score of a random predictor. This transformation measures improvement over baseline relative to the maximal achievable score, mapping baseline performance to 0 and perfect performance to 1. Source-level and room-level scores are obtained by normalizing each task score and averaging across the corresponding tasks.

\subsection{Sensitivity Analysis}

Probing measures how well spatial factors can be decoded from representations but does not reveal how embeddings respond to controlled perturbations. We measure representation sensitivity by comparing embeddings of scenes that differ only in source or room factors. Room sensitivity fixes the source configuration while varying room factors, whereas source sensitivity fixes the room configuration while varying source factors.

Let $f(x)\in\mathbb{R}^d$ denote the frozen embedding of input $x$. For a reference scene $x$ and paired variant $x'$ differing only in one factor group, we compute cosine similarity $s=\cos(f(x),f(x'))$. Because embedding spaces can exhibit different random-pair similarity levels across models, similarities are normalized relative to the expected similarity between unrelated samples. Sensitivity is defined as $\Delta(x,x') = 1 - (s-\mu)/(1-\mu)$, where $\mu=\mathbb{E}_{x_i,x_j}[\cos(f(x_i),f(x_j))]$ is the expected cosine similarity between random embeddings, estimated using 10{,}000 randomly sampled pairs from the test set.

Reference scenes $x$ are rendered from the test split. For each reference scene, a paired scene $x'$ is generated by modifying only the target factor group while keeping all other scene properties fixed. Sensitivity $\Delta$ is computed for each pair and averaged across all pairs.

\section{Results}

We analyze how spatial information is encoded in pretrained audio representations from four perspectives: input format, training paradigm, the source--room gap in probing performance, and representation sensitivity. We first examine how input format and training paradigm influence probing performance, and then analyze systematic differences between source and room factors in both decodability and geometric sensitivity.

\begin{figure}[!t]
\centering
\includegraphics[width=8.5cm]{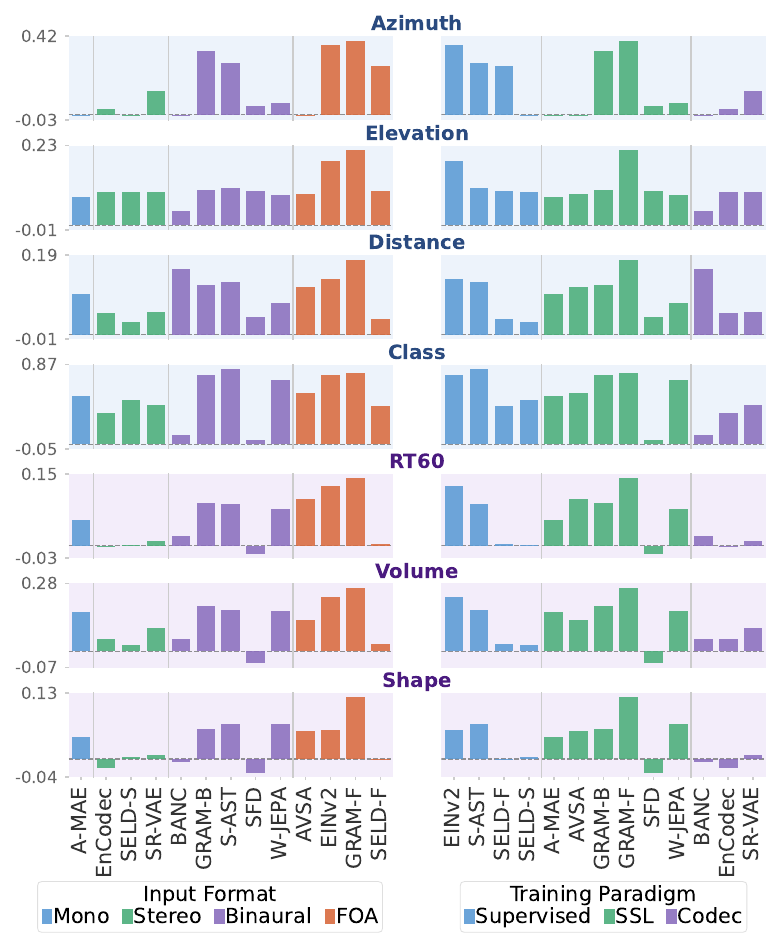}
\caption{Probing performance across spatial factors shown as improvement over a random predictor baseline. Rows correspond to prediction tasks (azimuth, elevation, distance, class, RT60, volume, shape). Models are ordered by input format (left) and by training paradigm (right).}
\label{fig:probing_results}
\end{figure}

\subsection{Input Format Effects}

The first column of Fig.~\ref{fig:probing_results} groups probing performance by input format (mono, stereo, binaural, FOA). Overall, representations derived from spatially structured multi-channel formats perform better across all tasks. In particular, binaural and FOA encoders consistently outperform mono and stereo models, indicating that richer spatial representations improve the encoding of spatial structure in the learned embeddings. Among the multi-channel formats, FOA models generally achieve the strongest performance, suggesting that the spherical harmonic representation provides a more informative spatial basis.

The magnitude of this advantage varies across tasks. For source-level factors such as azimuth, distance, and event classification, binaural and FOA models perform comparably, indicating that these factors are well captured by binaural spatial cues. In contrast, FOA models show advantages for elevation and for room-level properties such as RT60, volume, and shape. These results suggest that higher-order spatial representations provide additional information that is particularly beneficial for vertical localization and global room characteristics.

Despite lacking spatial channels, the mono A-MAE encoder performs competitively on all room-level tasks (RT60, volume, and shape) and most source tasks, failing primarily on azimuth estimation. This indicates that many spatial attributes can be inferred from monaural spectral–temporal structure. Reverberation and room geometry influence the decay profile and spectral characteristics of a signal, allowing models to recover aspects of the acoustic environment even from mono audio.

However, input format alone does not determine representational quality. Several binaural and FOA encoders perform poorly relative to other models with the same input format, indicating that access to spatial cues is not sufficient to guarantee strong spatial representations. This observation motivates a closer examination of how different training objectives influence spatial encoding, which we analyze next.

\subsection{Training Paradigm Effects}

The second column of Fig.~\ref{fig:probing_results} groups models by training paradigm (self-supervised, supervised, and codec-based), following the taxonomy in Table~\ref{tab:spatial_models}. Clear differences emerge across paradigms. Supervised localization models achieve the strongest performance on azimuth, reflecting their explicit training for sound event detection and direction-of-arrival estimation. However, this advantage does not extend to room-level factors: supervised models, particularly SELD-S and SELD-F, perform consistently poorly on RT60, volume, and shape. This suggests that localization supervision encourages representations that emphasize directional cues while discarding information about global acoustic context.

In contrast, self-supervised learning produces more balanced spatial representations. Several SSL encoders achieve more consistent performance across both source and room tasks, indicating that reconstruction- or prediction-based objectives can retain a broader range of spatial cues without explicit task supervision. Codec-based models show the weakest overall performance, suggesting that compression objectives optimized for perceptual fidelity and bitrate efficiency do not preserve spatial structure in a form that is easily decodable by probing.

Within the SSL paradigm, objective design plays a critical role. GRAM-B and GRAM-F, which reconstruct masked spectrogram patches directly in the input space, consistently achieve strong probing performance. In contrast, W-JEPA predicts latent representations and SFD distills derived spatial features (e.g., interaural differences), both producing weaker or less stable spatial encoding. These results suggest that reconstruction applied directly to spatially structured inputs preserves source and room information more effectively than objectives operating at more abstract representational levels.

\subsection{Source--Room Performance Gap}

\begin{figure}
\centering
\includegraphics[width=8.5cm]{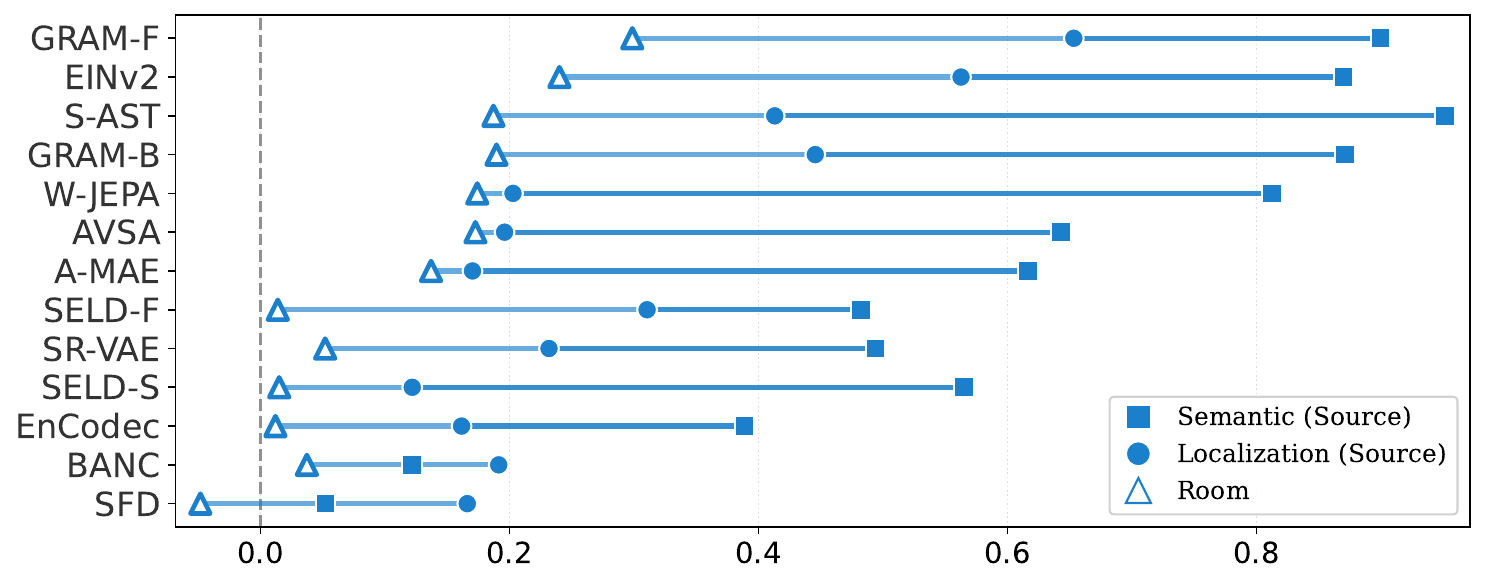}
\caption{Aggregated probing performance across factor groups after baseline normalization. 
Squares denote semantic source tasks (class), circles denote localization source tasks 
(azimuth, elevation, distance), and triangles denote room tasks (RT60, volume and shape).}
\label{fig:temp}
\end{figure}

Fig.~\ref{fig:temp} summarizes normalized improvement over the random baseline for three factor groups. Localization improvement is averaged across azimuth, elevation, and distance, while room improvement is averaged across RT60, volume, and shape. Event classification is reported separately as a semantic factor. Across all encoders, source-related factors exceed room factors: both localization and semantic improvements are larger for every model, revealing a systematic gap between source and room information in pretrained audio representations.

The two source components exhibit different patterns. Semantic improvement is uniformly strong across models, with most encoders achieving large gains above the baseline. Localization improvement is generally strong but shows greater variability across models, indicating that directional cues are captured with varying effectiveness depending on model design. In contrast, room improvement remains consistently smaller and often closer to the baseline, suggesting that global room properties are harder to recover from the learned representations.

A structural explanation may underlie this pattern. Source factors are localized and directly observable from a single recording, whereas room factors reflect global properties of the acoustic environment. A waveform implicitly contains the room impulse response for one source–receiver configuration, but inferring room geometry or volume would require observations across multiple source and listener positions. Consequently, typical pretraining setups provide stronger signals for encoding source-related information than  global room properties.

\subsection{Representation Sensitivity}

\begin{figure}[!t]
\centering
\includegraphics[width=8.5cm]{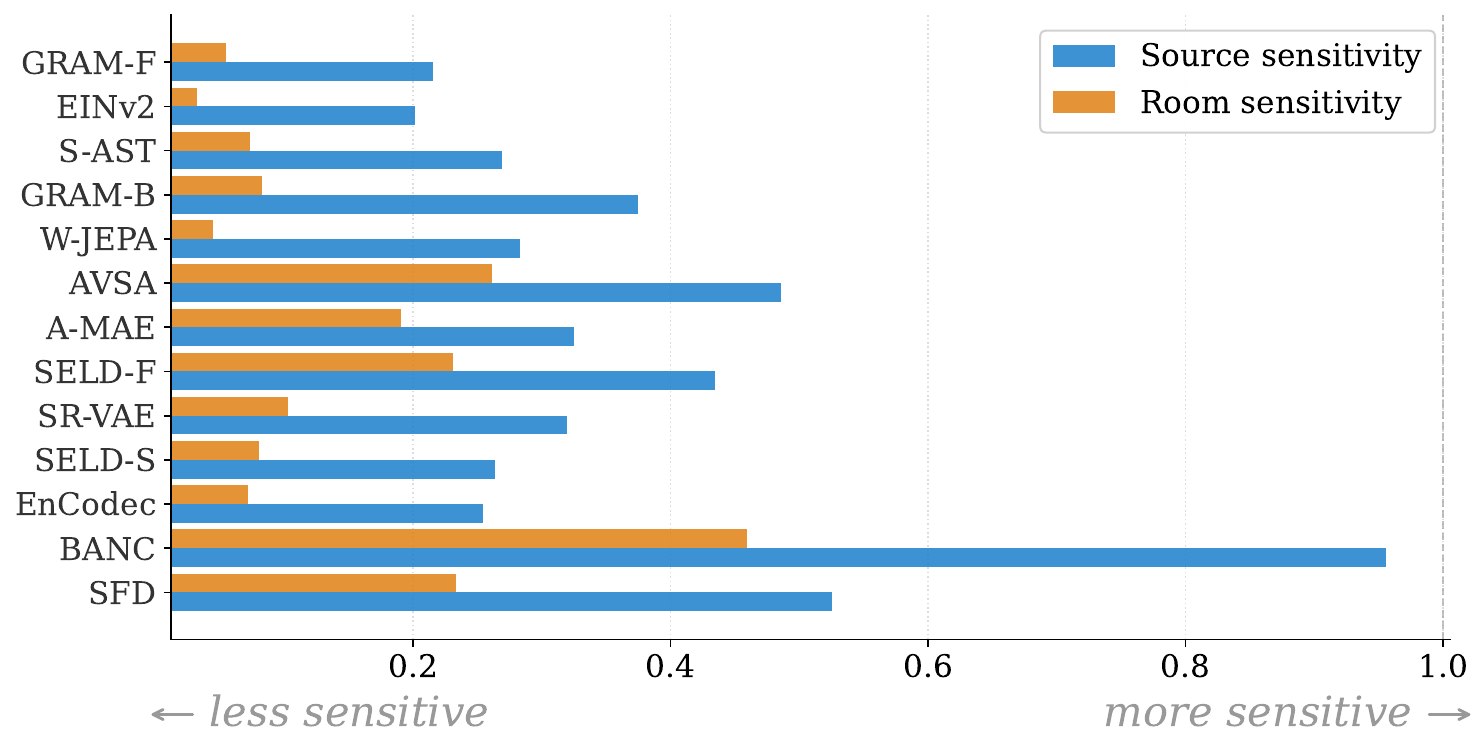}
\caption{Representation sensitivity to controlled perturbations of source and room factors. 
Sensitivity is measured as $\Delta(x,x')$ between paired scenes differing only in the selected factor group. Bars show source sensitivity and room sensitivity for each model.}
\label{fig:embedding_invariance}
\end{figure}

Fig.~\ref{fig:embedding_invariance} summarizes normalized representation sensitivity to controlled source and room perturbations, with models ordered according to aggregated probing improvement (Fig.~\ref{fig:temp}). Across nearly all encoders, perturbations of source factors produce larger embedding changes than room perturbations, indicating that pretrained representations are generally more responsive to source-level variation than to changes in global room properties.

This pattern is consistent with the probing results. Factors that are easier to decode in probing also tend to induce larger representation changes, whereas room perturbations typically produce smaller shifts. Sensitivity therefore provides a geometric counterpart to the source–room gap observed in probing.

Sensitivity magnitudes vary substantially across models and do not track probing performance monotonically. Some of the strongest models, such as GRAM-F and EINv2, exhibit relatively low overall sensitivity, whereas weaker models such as SFD and BANC produce much larger embedding shifts under both source and room perturbations. Models with intermediate probing performance span a broad range of sensitivities, indicating that larger representation changes do not necessarily correspond to better decodability. Instead, these results suggest that stronger models encode spatial factors in a more stable and structured manner, rather than via large embedding fluctuations.

\section{Conclusion}

We introduced a controlled framework for evaluating spatial factor encoding in pretrained audio representations. The study combines a synthetic dataset with independently controllable spatial factors, a unified probing benchmark spanning seven source and room tasks, and a complementary representation sensitivity analysis that measures embedding responses under controlled perturbations.

Our results reveal a consistent performance gap between source and room factors in learned representations. Across models, spatial attributes related to sound sources are substantially more accessible than global room properties. This pattern appears both in probing performance and in representation sensitivity, where embeddings exhibit larger deviations under controlled source variation than under room variation. Models trained with spatial inputs, particularly FOA, together with self-supervised learning objectives tend to produce the most robust spatial factor encoding.

Our analysis is conducted in a controlled synthetic environment with single-source scenes, which simplifies real-world acoustic complexity. Evaluation relies on linear probes that measure linearly accessible information in frozen embeddings, using mean-pooled representations to enable architecture-agnostic comparison despite differences in temporal and spectral resolution across models; examining pre-pooled features remains an important direction for future work. Models are also tested under a distribution that differs from their original training conditions. Extending the benchmark to real recordings, multi-source environments, and alternative probing strategies remains an important direction. The proposed framework can also serve as a diagnostic tool for developing spatially-aware audio representation models.

\newpage

\section{Acknowledgments}

This work is partially funded by the NYU / SONY Audio Institute for Music Business and Technology.

\section{Generative AI Use Disclosure}

Generative AI tools were used only for limited language editing and polishing. All scientific ideas, experiments, source-code, and conclusions were developed and verified by the authors, who take full responsibility for the manuscript.

\bibliographystyle{IEEEtran}
\bibliography{mybib}

\end{document}